\documentclass[conference]{IEEEtran}

  	\usepackage[pdftex]{graphicx}
  	\graphicspath{{../pdf/}{../jpeg/}}
	\DeclareGraphicsExtensions{.pdf,.jpeg,.png}

	\usepackage[cmex10]{amsmath}
	\usepackage{mathabx}
	\usepackage{algorithmic}
	\usepackage{array}
	\usepackage{mdwmath}
	\usepackage{mdwtab}
	\usepackage{eqparbox}
	\usepackage{url}
    \usepackage{xcolor}
    \usepackage{cite}  
    \usepackage{multirow}
    \usepackage{url}  
    \usepackage{breakurl}  

\begin{document}

\title{\LARGE Bushfire Severity Modelling and Future Trend Prediction Across Australia: Integrating Remote Sensing and Machine Learning}


\author{\authorblockN{Shouthiri Partheepan\authorrefmark{1}\authorrefmark{2}, Farzad Sanati \authorrefmark{1}, Jahan Hassan\authorrefmark{1} }
 \authorblockA{\authorrefmark{1}School of Engineering and Technology, Central Queensland University, \\ Norman Gardens, QLD 4701, Australia}
 \authorblockA{\authorrefmark{2}Department of Computing, Faculty of Science, Eastern University Sri Lanka, \\ Vantharumoolai, Chenkalady,
Batticaloa 30000, Sri Lanka}}

\maketitle

\begin{abstract}

Bushfire is one of the major natural disasters that cause huge losses to livelihoods and the environment. Understanding and analyzing the severity of bushfires is crucial for effective management and mitigation strategies, helping to prevent the extensive damage and loss caused by these natural disasters. This study presents an in-depth analysis of bushfire severity in Australia over the last twelve years, combining remote sensing data and machine learning techniques to predict future fire trends. By utilizing Landsat imagery and integrating spectral indices like NDVI, NBR, and Burn Index, along with topographical and climatic factors, we developed a robust predictive model using XGBoost. The model achieved high accuracy, 86.13\%, demonstrating its effectiveness in predicting fire severity across diverse Australian ecosystems. By analyzing historical trends and integrating factors such as population density and vegetation cover, we identify areas at high risk of future severe bushfires. Additionally, this research identifies key regions at risk, providing data-driven recommendations for targeted firefighting efforts. The findings contribute valuable insights into fire management strategies, enhancing resilience to future fire events in Australia. Also, we propose future work on developing a UAV-based swarm coordination model to enhance fire prediction in real-time and firefighting capabilities in the most vulnerable regions.

\end{abstract}

\IEEEoverridecommandlockouts

\IEEEpeerreviewmaketitle


\section{Introduction}

Bushfires are a natural and recurring phenomenon in many parts of the world, but they hold particular significance in Australia due to the continent’s unique climatic conditions, vegetation types, and land management practices \cite{ellis2004national}. Historically, fire has played an essential role in shaping Australian ecosystems, aiding in the regeneration of native plant species and maintaining biodiversity \cite{clarke2021fire}. However, the increasing frequency and severity of bushfires over recent years have raised critical concerns regarding the underlying causes and the potential long-term impacts on both natural and human systems.

Bushfire occurrences have increased significantly in numerous regions around the world, causing serious concern \cite{makumbura2024spatial}. Particularly, the devastating 2019-2020 Black Summer bushfires in Australia are a reminder of how catastrophic these events can be. Over 18.6 million hectares were burnt, resulting in the loss of 33 human lives, the destruction of over 3,500 homes, and an estimated one billion animals killed \cite{recovery2020}. Such events highlight the immediate dangers of bushfires and underscore the need for comprehensive strategies to manage and mitigate their impacts. By combining remote sensing with geographic information systems (GIS), bushfire risk zones can be accurately mapped \cite{feizizadeh2023integrated}, which is crucial for managing bushfire crises. Accurate mapping and analysis of fire severity are vital for developing these strategies and providing essential information for policymakers, land managers, and emergency services.

The relationship between climate change and bushfire activity has been extensively studied \cite{collins2018utility, collins2020training, dixon2022regional, gai2011gis}, with numerous researchers highlighting the increasing risk of severe fire events due to rising temperatures and prolonged droughts. Dowdy et al. \cite{dowdy2018climatological} analyzed climate models to predict fire weather conditions, indicating that higher temperatures and reduced rainfall are likely to increase the frequency and intensity of bushfires. Similarly, Nolan et al. \cite{nolan2020causes} assessed the impact of fuel load and dryness on fire behaviour, emphasizing the need for adaptive management strategies to cope with changing fire regimes. Vigilante et al. \cite{vigilante2024factors} highlight the effectiveness of integrating traditional Indigenous fire management practices with modern technologies in managing fire regimes. They demonstrate through the work of the North Kimberley Fire Abatement Project that strategic early dry-season burning can reduce the frequency of severe late dry-season wildfires. This model could be adapted to improve fire severity prediction models in other regions.

Recent studies on fire severity in Australia have employed advanced remote sensing techniques along with machine learning models to produce more accurate and efficient fire severity mapping. Machine learning and UAV (Unmanned Aerial Vehicle) technology improve predictive capabilities by reducing firefighter response times and offer effective tools for real-time forest fire detection and monitoring \cite{partheepan2023a}. According to our previous study \cite{partheepan2023b}, pre-trained deep-learning models outperformed custom models in early bushfire detection. We illustrate the importance of leveraging pre-trained models for improving fire detection accuracy, offering a pathway for integrating these models with fire severity predictions. Collins et al. \cite{collins2018utility} utilized Random Forest (RF) classifiers within the Google Earth Engine (GEE) platform to rapidly produce fire severity maps, demonstrating the method's potential for large-scale application and continuous improvement over time with new data. Similarly, Dixon et al. \cite{dixon2022regional} applied a machine learning framework to the Northern Jarrah Forest region, combining Landsat-derived spectral indices and high-resolution post-fire aerial imagery to produce a 16-year history of fire severity. This model achieved high accuracy and provided valuable insights into wildfire's spatial and temporal patterns and prescribed burn severity. 

Understanding the trends and predicting future bushfire occurrences are crucial for developing proactive management strategies. In this study, we provide detailed spatial analysis and trend predictions, contributing to the existing body of knowledge. Furthermore, it addresses the critical gap of comprehensive, data-driven strategies for reducing the adverse effects of bushfires, thereby enhancing Australia’s resilience to future fire events. 

The primary objectives of this paper are: 
\begin{itemize}
\item Mapping fire severity across Australia over the past twelve years using NASA's FIRMS(Fire Information for Resource Management System) satellite data.
\item Identifying and analysing trends and patterns in fire occurrence and severity.
\item Developing predictive models for future fire trends based on historical data.
\item Offering recommendations for targeted firefighting efforts, considering population density and vegetation types.
\end{itemize}

This paper is structured to provide a comprehensive approach to the study, starting with Section 2, which outlines the methodology. In this section, we detail the process of selecting the study area and data collection methods and elaborate on the prediction model, where we discuss the machine learning approach used to forecast fire severity. Additionally, we highlight the influential factors that guided our recommendations for fire management and resource allocation, linking these factors to the predictive outcomes. The subsequent section presents the results of the study, incorporating detailed analyses supported by appropriate visualizations. Finally, the paper concludes with a reflection on the key findings and their implications for fire management practices. 


\section{Methodology}

\subsection{Study Area}

Australia (25.2744° S, 133.7751° E) spans a vast and diverse landmass of approximately 7.692 million square kilometres, making it the sixth-largest country in the world. Home to over 27 million people \cite{ABS2024}, Australia is renowned for its unique and varied ecosystems, which include tropical rainforests, eucalyptus woodlands, arid deserts, and alpine regions \cite{brown1997australian}. Australia's forest cover has significantly changed over the years due to natural and anthropogenic factors, including bushfires, logging, and land clearing for agriculture and urban development \cite{ramankutty2006global}.

\begin{figure}[ht!] 
\centering
\includegraphics[width=3.5in]{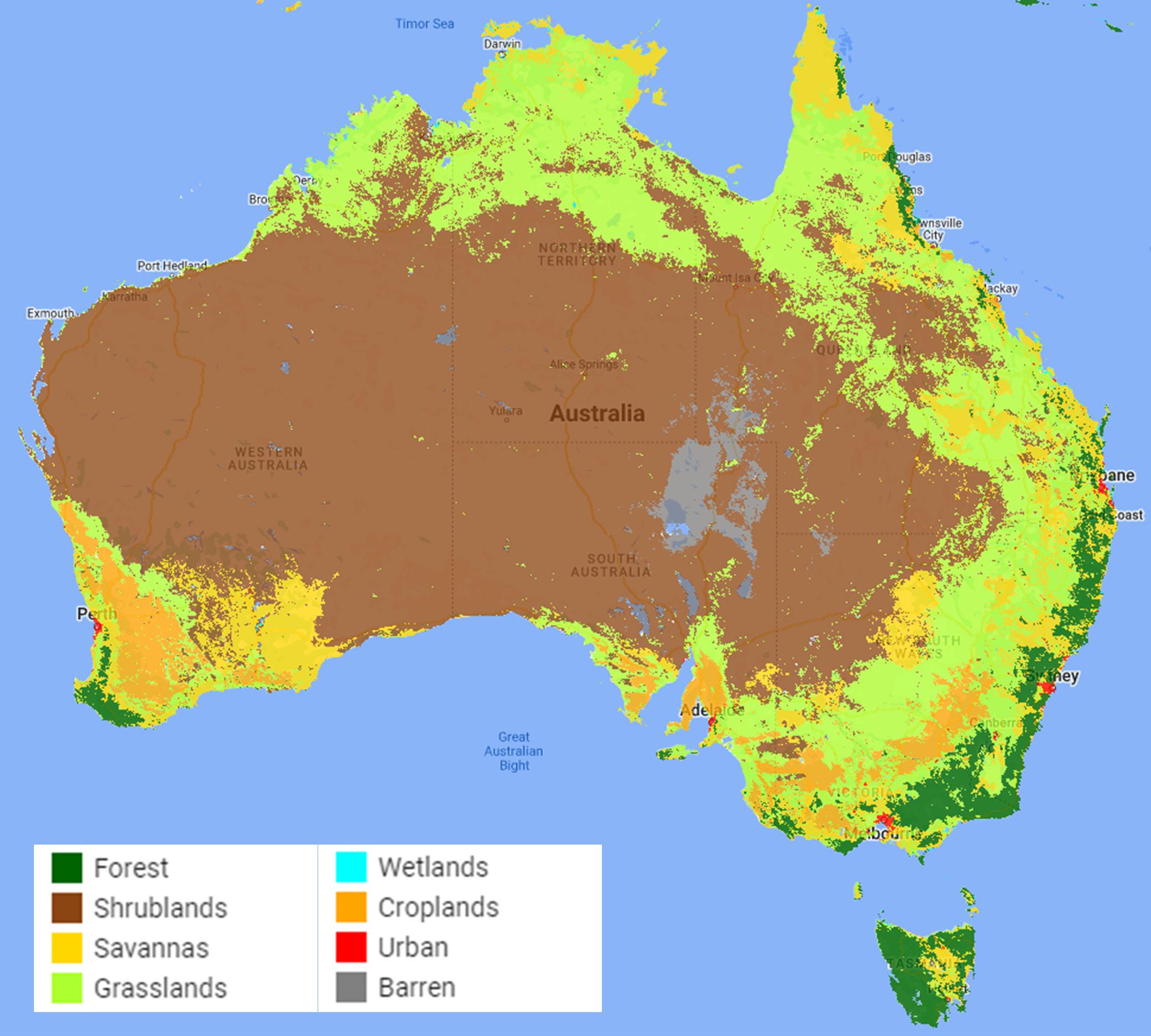}
\caption{Map of Australia's Land Cover Types in 2018 on QGIS: This map visualizes using QGIS and visualizes various land cover types across Australia, categorized into Forests, Shrublands, Savannas, Grasslands, Wetlands, Croplands, Urban areas, and Barren lands. The data is derived from the MODIS Land Cover Type Yearly Global 500m dataset. Forest areas are marked in dark green, Shrublands in brown, Savannas in yellow, Grasslands in light green, Wetlands in cyan, Croplands in orange, Urban areas in red, and Barren lands in grey. The map provides a clear overview of the dominant land cover types throughout the country.}
\label{figure1}
\end{figure}

The study area encompasses the entirety of Australia, a continent characterized by diverse ecosystems, from tropical rainforests in the north to temperate forests in the south and arid deserts in the interior. Fig.\ref{figure1} provides a detailed representation of the land cover distribution across Australia, highlighting the various vegetation types, urban areas, and other key landscape features. The land cover map (Fig. \ref{figure1}) was created using raster data from the Australian Government Department of Agriculture, Fisheries and Forestry (2023). This dataset offers detailed spatial data on forest cover across Australia.

Australia has recurrent and significant bushfires every year, which are heavily influenced by the country's climate and vegetation. The combination of hot, dry summers and highly flammable eucalyptus forests creates ideal conditions for bushfires. These ecosystems vary significantly in their fire regimes, with some adapted to frequent low-intensity fires and others to infrequent but severe fires. This comprehensive analysis is crucial in enhancing our understanding of fire behaviour in different Australian ecosystems and can potentially revolutionize fire management practices.

\subsection{Fire Trend Analysis}
\subsubsection{Data Collection}

MODIS (Moderate Resolution Imaging Spectroradiometer) collects data across multiple time frames with sophisticated remote sensing capabilities \cite{xiong2009overview}. MODIS provides data at various spatial resolutions, including 250 m, 500 m, and 1 km, which allows for flexibility depending on the scale of the analysis required \cite{giglio2003enhanced}. This versatility in spatial resolution helps enhance the accuracy of monitoring fire events and their severity across different landscapes. During the fire season, MODIS offers the potential to improve bushfire management by providing near-real-time data for monitoring active fires and burnt areas \cite{chuvieco2020satellite}. As a key sensor of NASA's Earth Observation System (EOS), MODIS delivers critical data through its Terra and Aqua satellites, offering twice-daily coverage of most locations worldwide, thus significantly enhancing bushfire monitoring and control efforts.

This study used data spanning 12 years, from February 2012 to January 2024, obtained from the NASA FIRMS (https://firms.modaps.eosdis.nasa.gov/). This extensive dataset encompasses fire events detected across various regions of Australia, focusing on understanding fire severity and predicting future fire trends. The primary purpose is to map fire severity and provide insights on areas needing more attention based on factors like population and vegetation.

The detection confidence index recognizes the level of confidence in detecting fire and can range from 0\% to 100\%. Pixels are assigned to either low-confidence fire, nominal-confidence fire, or high-confidence fire based on the Detection confidence. This classification process yields better accuracy when the threshold is above 30\%, as suggested by Giglio et al. \cite{giglio2003enhanced} in their previous research.

\subsubsection{Data Preprocessing}
In this study, data preprocessing was a crucial step to ensure the accuracy and reliability of the fire severity mapping and trend analysis. The preprocessing involved multiple steps in cleaning, normalizing, and preparing the dataset for visualization and analysis.

First, all duplicate values were identified and removed from the dataset to maintain the integrity of the data and ensure that each fire event was counted only once. Next, normalization was applied to the dataset. Given the various attributes with different scales, normalization standardized the data attributes to a common range, typically between 0 and 1. The normalization process improved the convergence rate of the learning algorithms and ensured that no single attribute disproportionately influenced the model. For detailed visualization and analysis, the dataset was split into yearly subsets for a comprehensive examination of fire severity trends and patterns over the past twelve years. Analyzing the data on a yearly basis enabled the observation of annual variations in fire occurrences and severity, facilitating a more granular understanding of the temporal dynamics of bushfires in Australia.

Any missing values in the dataset were addressed to prevent potential biases in the analysis. Finally, the dataset underwent thorough validation and consistency checks to verify the accuracy of the cleaned data. This step involved cross-referencing with original data sources and ensuring that the preprocessing steps did not introduce any errors or inconsistencies. The comprehensive preprocessing of the dataset ensured that the data used for fire severity mapping and trend analysis was accurate, reliable, and suitable for detailed visualization and modelling. 

We utilized QGIS \cite{QGIS2024}, an open-source Geographic Information System (GIS) platform, to visualize and map fire severity trends over the past 12 years. QGIS offers robust tools for handling various geospatial data formats, including shapefiles, making it an ideal choice for our analysis \cite{vitalis2020cityjson}. By leveraging QGIS, we were able to effectively visualize historical fire severity data, allowing us to identify patterns and trends in fire activity across different regions. This platform's flexibility and extensive functionality were instrumental in our comprehensive analysis of fire severity over time.

\subsection{Predictive Model }
\begin{figure}[ht!] 
\centering
\includegraphics[width=3.5in]{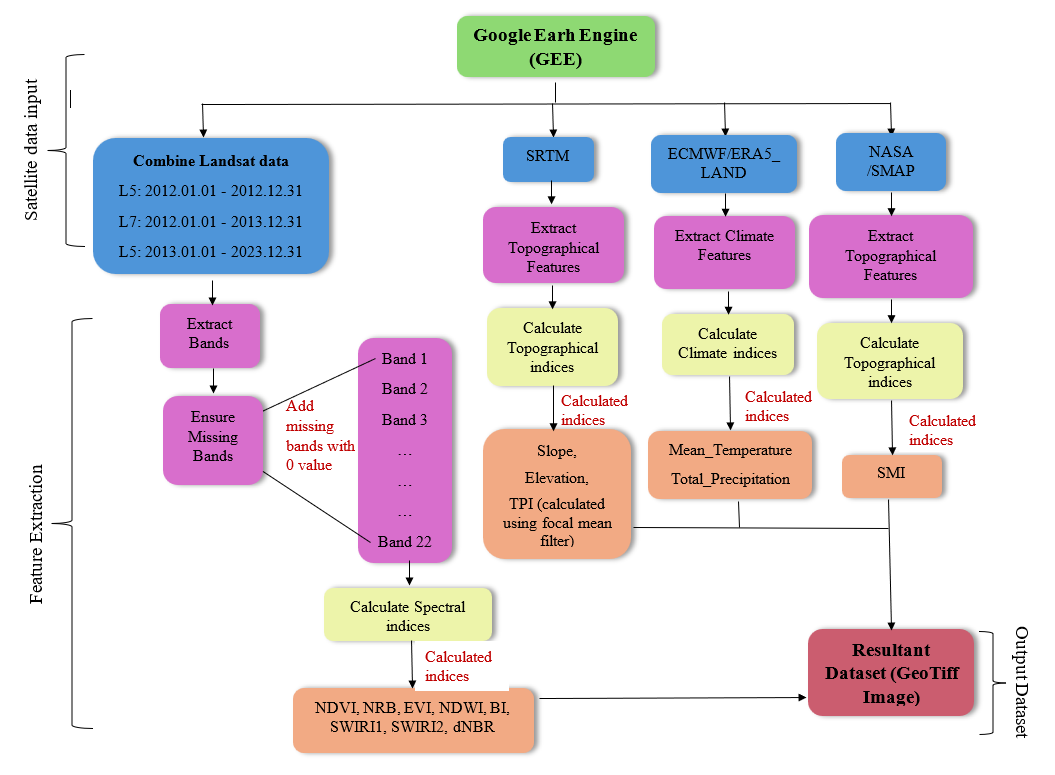}
\caption{Data extraction workflow for fire severity analysis using multi-source satellite imagery. The process involves combining Landsat data and additional sources such as SRTM, ECMWF/ERA5-Land, and NASA/SMAP to extract topographical, climatic, and spectral indices. Missing bands are handled with zero values, and the calculated indices are used to produce a GeoTIFF output dataset.}
\label{figure2}
\end{figure}

The predictive model is crucial in this study as it anticipates future fire severity trends across Australia, informing strategic resource allocation and risk mitigation efforts in fire-prone regions. Remote sensing, using satellite and aerial imagery, is particularly advantageous for fire severity mapping due to its ability to cover large areas and provide consistent, repeatable observations. Fig \ref{figure2} provides a comprehensive overview of the entire process involved in feature extraction from multiple satellite imagery sources, culminating in the creation of the final dataset. The subsequent sections delve into the detailed steps of the data extraction process, outlining each stage, from raw data acquisition to the calculation of the relevant indices.

\subsubsection{Data Collection }
The data collection process utilized Landsat imagery from Landsat 5, 7, and 8 satellites to cover the period from 2012 to 2023. The region of interest (ROI) was defined as Australia, using specific geographical coordinates \cite{AustraliaROI2024} to encompass the entire country. Data was acquired using the Google Earth Engine (GEE) platform, which facilitates cloud-based image processing and large-scale analysis.

From 2013 to 2023, Landsat 8 data was used, while Landsat 5 and Landsat 7 were employed for earlier years (2012 to 2013). Each satellite offers different bands, and to ensure consistency across datasets, missing bands in Landsat 5 and 7 were filled with constant values to ensure a standard set of 22 bands across all images. We utilized advanced processing techniques to reduce atmospheric interference and enhance the reliability of the data, particularly by working with the cloud cover and quality bands.

The standard set of 22 bands included surface reflectance bands (e.g., SR\_B1, SR\_B2), thermal bands (ST\_B6, ST\_B10), and QA (quality assessment) bands. Each image was processed to standardize this band order and ensure uniformity across all datasets. The data was extracted at a 500-meter scale, balancing spatial resolution and computational efficiency to provide detailed yet manageable datasets for further analysis. The detailed band values extracted from the Landsat are listed in Table \ref{tab:band_info}. This table presents the 22 bands extracted from Landsat 5, 7, and 8 satellite imagery, including surface reflectance, thermal, and quality assessment bands. The minimum, maximum, and mean values for each band are also provided to illustrate the range of data used in the analysis. 

\begin{table*}[h!]
\centering
\caption{Summary of Extracted Bands from Landsat Imagery with Descriptions, Maximum, Minimum, and Mean Values}
\resizebox{\textwidth}{!}{%
\begin{tabular}{|l|c|c|c|c|c|}
\hline
\textbf{Band Number}      & \textbf{Band Name} & \textbf{Description} & \textbf{Maximum Value} & \textbf{Mean value} & \textbf{Minimum Value} \\ \hline
\textbf{Band 1} & $SR\_B1$    & Surface Reflectance (Ultra Blue - L8 and Blue - L5, L7)     & 65535     & 8657.67       & 1191                                     \\ \hline
\textbf{Band 2} & $SR\_B2$    & Surface Reflectance (Blue -L8 and Green - L5, L7)          & 65535     & 9050.00       & 1174                                    \\ \hline
\textbf{Band 3} & $SR\_B3$    & Surface Reflectance (Green -L8 and Red - L5, L7)           & 65535     & 10134.42      & 2142                                          \\ \hline
\textbf{Band 4} & $SR\_B4$    & Surface Reflectance (Red -L8 and Near Infrared - L5, L7)   & 44845     & 11436.10      & 2504                                        \\ \hline
\textbf{Band 5} & $SR\_B5$    & Surface Reflectance (Near Infrared -L8 and Shortwave Infrared 1 - L5, L7)  & 42836     & 14204.63   & 5683                                        \\ \hline
\textbf{Band 6} & $SR\_B6$    & Surface Reflectance (Shortwave Infrared 1 - L8)            & 29703.75      & 13929.94  & 0                                      \\ \hline
\textbf{Band 7} & $SR\_B7$    & Surface Reflectance (Shortwave Infrared 2)                & 27255.5       & 13804.80  & 7236                                          \\ \hline
\textbf{Band 8} & $SR\_QA\_AEROSOL$ & Aerosol Quality Assessment - L8                      & 228           & 91.27     & 0                                        \\ \hline
\textbf{Band 9} & $SR\_ATMOS\_OPACITY$  & Atmospheric Opacity – L5, L7                    & 0             & 0         & 0                                    \\ \hline
\textbf{Band 10}    & $SR\_CLOUD_QA$   & Cloud Quality Assessment – L5, L7                & 0             & 0         & 0                                     \\ \hline
\textbf{Band 11}    & $ST\_B6$    & Surface Temperature – L5, L7                          & 0             & 0         & 0                                          \\ \hline
\textbf{Band 12}    & $ST\_B10$   & Surface Temperature - L8                              & 51973         & 39908.41  & 0                                        \\ \hline
\textbf{Band 13}    & $ST\_ATRAN$  & Atmospheric Transmission                & 9375.25       & 7357.86   & 3132                                       \\ \hline
\textbf{Band 14}    & $ST\_CDIST$  & Distance to Cloud                     & 21526         & 1678.99   & 0                                          \\ \hline
\textbf{Band 15}    & $ST\_DRAD$   & Downward Radiation                    & 2239          & 962.89    & 197.5                                        \\ \hline
\textbf{Band 16}    & $ST\_EMIS$   & Emissivity                           & 9904          & 9674.91   & 8333                                        \\ \hline
\textbf{Band 17}    & $ST\_EMSD$   & Emissivity Standard Deviation               & 14128.25      & 63.77     & 0                                        \\ \hline
\textbf{Band 18}    & $ST\_QA$ & Uncertainty of the surface Temperature      & 4099.25       & 282.08    & 137                                        \\ \hline
\textbf{Band 19}    & $ST\_TRAD$   & Thermal Radiance                        & 11817.75      & 9322.59   & 0                                        \\ \hline
\textbf{Band 20}    & $ST\_URAD$   & Upward Radiation                       & 5375          & 2030.40   & 347.25                                        \\ \hline
\textbf{Band 21}    & $QA\_PIXEL$  & Pixel Quality                         & 56598         & 21782.26  & 5472                                        \\ \hline
\textbf{Band 22}    & $QA\_RADSAT$ & Radiometric Saturation                & 0             & 0         & 0                                      
\\ \hline

\end{tabular}%

}
\label{tab:band_info}
\end{table*}

\subsubsection{Feature Extraction}
From the Landsat imagery, the spectral indices (NDVI (Normalized Difference Vegetation Index), NBR (Normalized Burn Ratio), EVI (Enhanced Vegetation Index), NDWI (Normalized Difference Water Index), Burn Index (BI), SWIR1, SWIR2 and dNBR (Difference Normalized Burn Ratio)) were calculated (Table \ref{tab:spectral_indices}) to derive fire-related features. These indices are crucial for analyzing and predicting fire severity. Each of these indices was computed for every image to create a comprehensive dataset of fire-related features.

\begin{table*}[h!]
\centering
\caption{Spectral Indices and Band Information with Calculation Formulas and Statistical Values}
\resizebox{\textwidth}{!}{%
\renewcommand{\arraystretch}{1.5} 
\begin{tabular}{|l|c|c|c|c|c|}
\hline
\textbf{Band Number}      & \textbf{Band Name} & \textbf{Calculation} & \textbf{Maximum Value} & \textbf{Mean value} & \textbf{Minimum Value} \\ \hline
\textbf{Band 23}    & NDVI(Normalized Difference Vegetation Index) (Fig. \ref{figure3}) & $\frac{(SR\_B5 - SR\_B4)}{(SR\_B5 + SR\_B4)}$  \cite{rouse1974monitoring} & 0.4194 & 0.0865  & -0.4773                    \\ \hline
\textbf{Band 24}    & NBR (Normalized Burn Ratio)  (Fig. \ref{figure4})                 & $\frac{(SR\_B5 - SR\_B7)}{(SR\_B5 + SR\_B7)}$ \cite{key2006landscape}     & 0.4876 & 0.0177  & -0.2130                    \\ \hline
\textbf{Band 25}    & EVI (Enhanced Vegetation Index)               & $2.5 \times \frac{(SR\_B5 - SR\_B4)}{(SR\_B5 + 6 \times SR\_B4 - 7.5 \times SR\_B2 + 1)}$ \cite{huete2002overview} & 89.2105                        & 0.3650                         & -291.6667                                          \\ \hline
\textbf{Band 26}    & NDWI (Normalized Difference Water Index)          & $\frac{(SR\_B3 - SR\_B5)}{(SR\_B3 + SR\_B5)}$ \cite{mcfeeters1996use}                      & 0.6284 & -0.1349  & -0.4718                  \\ \hline
\textbf{Band 27}    & BI (Burn Index)                                 & $\frac{(SR\_B7 - SR\_B5)}{(SR\_B7 + SR\_B5)}$  \cite{chuvieco2006use}                        & 0.2130 & -0.0177  & -0.4876                  \\ \hline
\textbf{Band 28}    & SWIR1 (Shortwave Infrared 1)                    & SR\_B6       & 29703.75  & 13929.9403 & 0                                                             \\ \hline
\textbf{Band 29}    & SWIR2 (Shortwave Infrared 2)                    & SR\_B7       & 27255.5   & 13804.7989 & 7236                                          \\ \hline

\textbf{Band 36}    & dNBR (Difference Normalized Burn Ratio)  & $NBR_{\text{pre-fire}} - NBR_{\text{post-fire}}$  \cite{miller2007quantifying}     & 0.5409   & -0.0025 & -0.3917
\\ \hline

\end{tabular}%
}
\label{tab:spectral_indices}
\end{table*}

\begin{figure}[ht!] 
\centering
\includegraphics[width=3.5in]{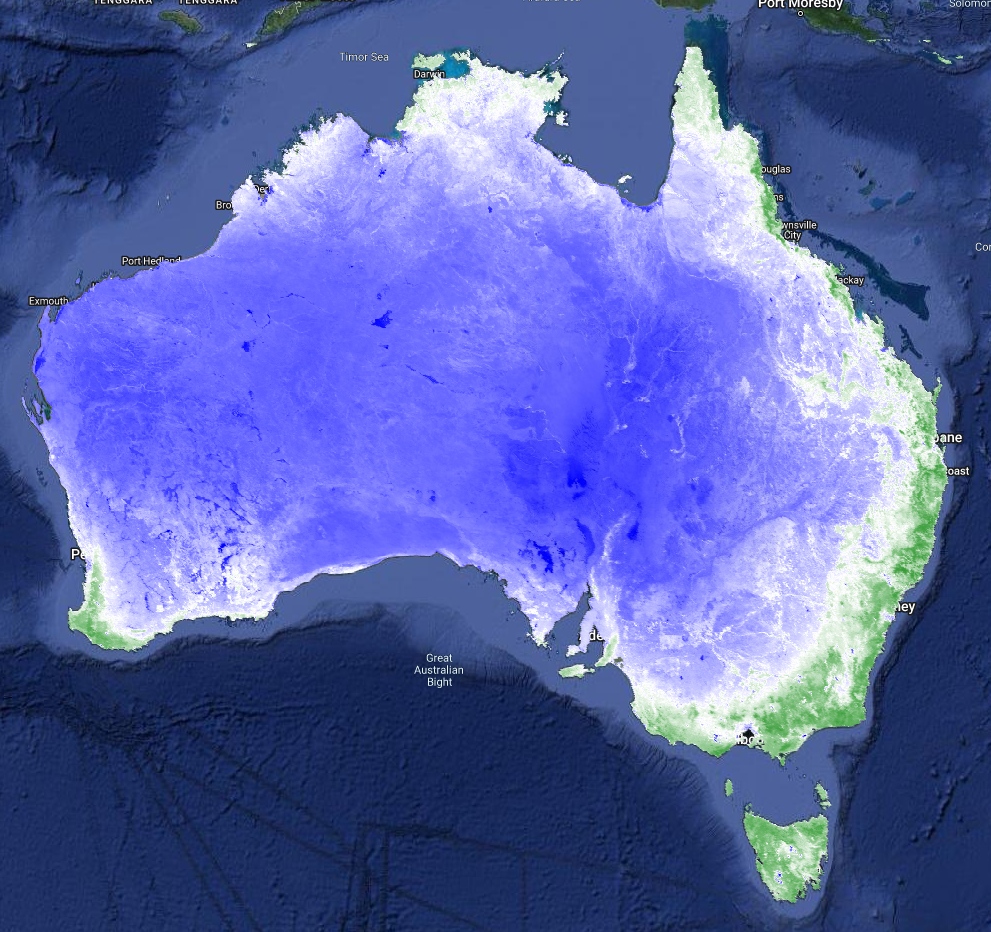}
\caption{NDVI Map of Australia (2012-2023) on GEE: This map depicts the vegetation health and density across Australia over a period of 12 years. Higher NDVI values, shown in green, indicate areas with dense and healthy vegetation, while lower values, depicted in blue, represent sparse or degraded vegetation cover.}
\label{figure3}
\end{figure}

\begin{figure}[ht!] 
\centering
\includegraphics[width=3.5in]{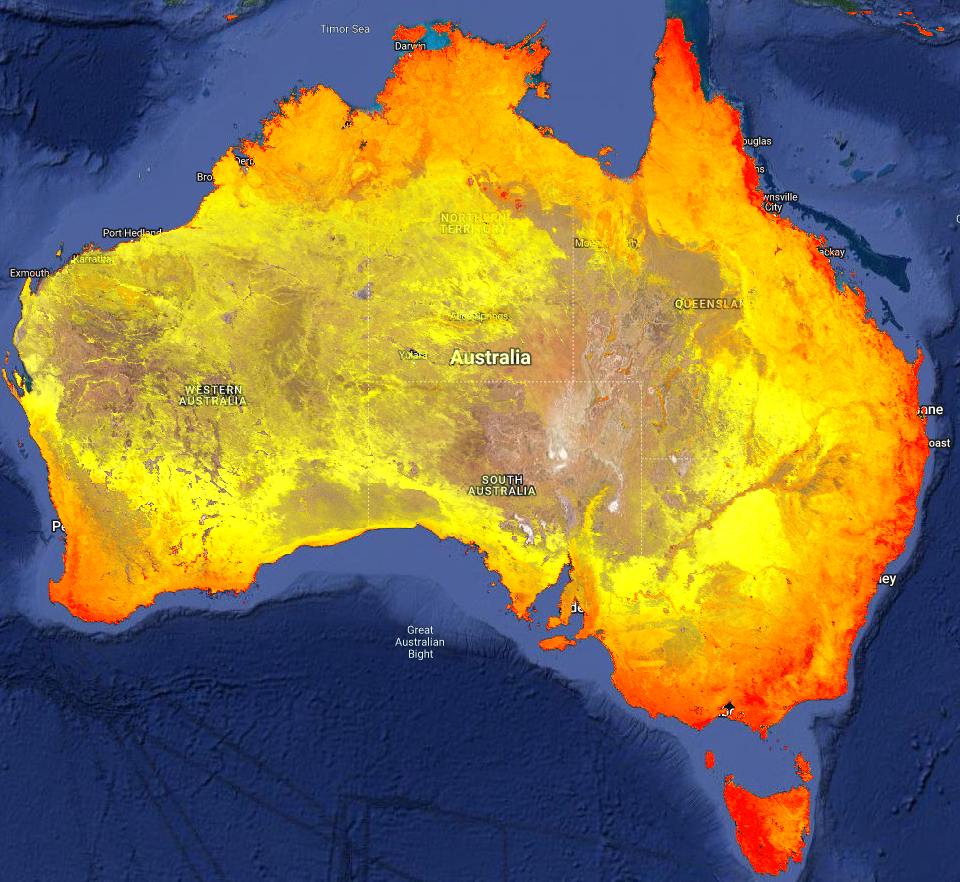}
\caption{NBR Map of Australia (2012-2023) on GEE: This map visualizes the burn severity across Australia using the NBR index, where warmer colors (yellow to red) indicate higher burn severity, and cooler colors (white) represent areas with lower or no burn impact over the last decade}
\label{figure4}
\end{figure}

The dNBR was calculated annually to assess fire severity over time. This index compares pre-fire and post-fire NBR values to evaluate the impact of fires on the landscape. The dNBR calculation was based on two defined periods for each year: the pre-fire period, from July 1 to September 30, and the post-fire period, from February 1 to March 30 of the following year. For each pixel, the NBR was computed for both the pre-fire and post-fire periods, with dNBR being the difference between these values. A positive dNBR indicates fire severity, with higher values reflecting more severe burn damage. This method measures the fire's impact on the landscape across different years.

\subsubsection{Topographical and Climatic Feature Extraction}
In addition to spectral indices, topographical and climatic features were incorporated to enhance the accuracy of fire severity predictions. Elevation and slope data were sourced from the Shuttle Radar Topography Mission (SRTM) dataset, where slope representing the elevation change rate was derived to understand better how terrain influences fire behaviour. The Topographic Position Index (TPI) was calculated as the difference between a pixel's elevation and the mean elevation of its surrounding area within a 500-meter radius, allowing for the identification of terrain features such as ridges and valleys, which play a critical role in fire spread.

Climatic factors were also considered, with mean temperature and total precipitation data extracted from the ECMWF ERA5 Land dataset. These variables provide vital context regarding the environmental conditions that influence fire occurrences. The Soil Moisture Index (SMI) was also derived from the NASA SMAP dataset, which measures moisture levels in the root zone. Low soil moisture can significantly increase fire risk, making SMI a key variable in predicting fire severity. This comprehensive inclusion of both topographical and climatic features creates a more robust model for assessing and predicting fire behaviour. The Table \ref{tab:other_features} presents the topographical and climatic features used in the analysis, including Elevation, Slope, TPI (Topographic Position Index), Mean Temperature, Total Precipitation, and Soil Moisture Index (SMI).

\begin{table*}[h!]
\centering
\caption{Topographical and Climatic Features with Maximum, Mean, and Minimum Values}
\resizebox{\textwidth}{!}{%
\begin{tabular}{|l|c|c|c|c|c|}
\hline
\textbf{Band Number}      & \textbf{Band Name} & \textbf{Description} & \textbf{Maximum Value} & \textbf{Mean value} & \textbf{Minimum Value} \\ \hline
\textbf{Band 30}    & Elevation  & Elevation from SRTM data & 2066 & 224.9916  & -15                    \\ \hline
\textbf{Band 31}    & Slope   & Slope derived from the elevation & 37.5421 & 0.7012  & 0                    \\ \hline
\textbf{Band 32}    & TPI (Topographic Position Index)  & TPI calculated using a focal mean filter with radius 500m & 175                       & 0.0022                        & -146.2000                                          \\ \hline
\textbf{Band 33}    & Mean\_Temperature          &   Mean temperature extracted from ECMWF ERA5-Land data & 302.7170 & 295.7664  & 279.4776                  \\ \hline
\textbf{Band 34}    & Total\_Precipitation               &     Total precipitation extracted from ECMWF ERA5-Land data   & 735.5974 & 69.5770  & 19.1339                  \\ \hline
\textbf{Band 35}    & SMI (Soil Moisture Index)                  &    SMI extracted from NASA SMAP root zone data    & 0.9083  & 0.1287 & 0.0189                                                            \\ \hline

\end{tabular}%
}
\label{tab:other_features}
\end{table*}

\subsubsection{Model Training and Evaluation}

In this study, we used the XGBoost regression model to predict fire severity, specifically using the dNBR as the target variable. XGBoost is a robust machine learning algorithm with gradient boosting to optimize predictive performance \cite{chen2016xgboost}. The implementation was done in Python 3.8.2, and the development environment used was Spyder IDE. In addition to XGBoost, various Python libraries, such as NumPy, pandas, and scikit-learn, were employed for data preprocessing, feature engineering, and model evaluation. The model was trained and tested on a High-Performance Computing (HPC) cluster with 100 GB of RAM and access to 20 CPUs. The exact resource utilization for this execution was not fully tracked, but given the large memory and CPU allocation, the system was able to handle the substantial dataset efficiently without memory bottlenecks. The HPC environment was beneficial for handling the computational demands of the XGBoost model, which required significant memory and processing power to manage the large feature set and extensive hyperparameter tuning process.

The dataset was extracted from a GeoTIFF file with 36 band values, with dNBR values stored in the last band and all other bands as features and reshaped accordingly. Feature scaling was applied using StandardScaler to standardize the inputs, enhancing the model's stability and performance. Our study was conducted thoroughly, as we manually tuned the model's hyperparameters by experimenting with various combinations to find the optimal configuration. After multiple trials, we identified the best hyperparameters: 500 boosting rounds (n\_estimators), a learning rate of 0.1, a maximum tree depth of 5, an 80\% subsampling rate for the training data, and 80\% feature sampling for each tree. These parameters struck a good balance between model complexity and generalization, enabling the model to capture significant patterns in the data without overfitting.

The dataset was split into training and testing sets, with 80\% of the data allocated for training and 20\% for testing. We performed 5-fold cross-validation on the training data during training to evaluate the model's generalization ability. The R-squared (R²) scores from cross-validation were used to measure the model's performance, with the mean R² score serving as a benchmark for its predictive capacity. After training the model, it was evaluated on the testing set using two key metrics: Mean Squared Error (MSE) and R-squared (R²). The MSE calculated the average squared difference between the dNBR value and the predicted value, whereas the R² demonstrated the predicted variance of dNBR. These evaluation metrics confirmed that the chosen hyperparameters provided robust performance on unseen data, validating the manual tuning process.

\subsection{Factors Influencing Firefighting Resource Allocation}
\subsubsection{Population Density and Urban Proximity}
Population density and proximity to urban areas are critical factors when allocating firefighting resources. High-population areas present a greater risk to human life and infrastructure, necessitating a more significant allocation of resources. According to the Australian Bureau of Statistics, over 86\% of Australians live in urban areas, with significant populations concentrated along the eastern coast \cite{ABS2024}. Regions such as New South Wales (NSW) and Victoria, where Sydney and Melbourne are located, have the highest population densities. During the 2019-2020 bushfire season, these areas faced significant threats, mobilising substantial firefighting resources to protect lives and properties. Firefighting efforts were concentrated in these high-risk areas, underscoring the importance of population density in resource allocation.

\begin{figure}[ht!] 
\centering
\includegraphics[width=3.5in]{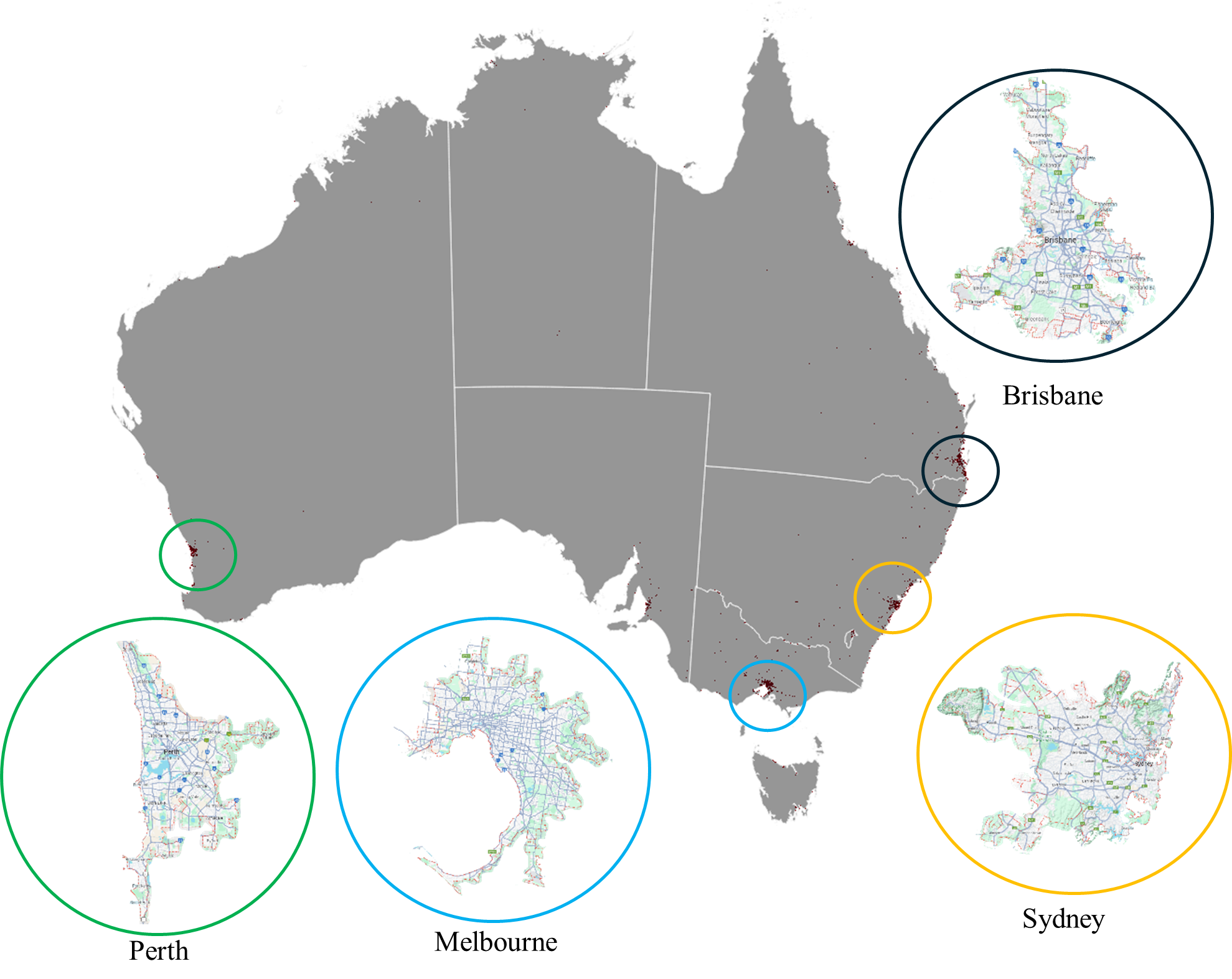}
\caption{The distribution of major urban centres across Australia (The dataset was acquired from \cite{worldpop_australia_2020} and visualized using QGIS), with particular attention paid to Sydney, Melbourne, Brisbane, and Perth. These areas are critical in firefighting resource allocation due to their high population densities and proximity to bushfire-prone regions.}
\label{figure5}
\end{figure}

Fig.\ref{figure5} highlights the major urban centres of Australia, including Sydney, Melbourne, Brisbane, and Perth, where population density is highest. The concentration of people in these urban areas, particularly along the eastern coast, increases the risk to human life and infrastructure during bushfire events. For example, Sydney and Melbourne, located in New South Wales and Victoria, respectively, are among the most densely populated areas and were severely impacted during the 2019-2020 bushfire season. The proximity of these urban centres to fire-prone regions necessitates a strategic and substantial allocation of firefighting resources to protect both lives and property. This map visually supports the argument that population density and urban proximity are vital considerations in resource distribution during fire emergencies.

\subsubsection{Vegetation and Fuel Load}
Vegetation type and fuel load significantly influence fire behaviour \cite{trollope1984fire} and severity. Australia’s diverse ecosystems range from temperate forests to grasslands and arid scrublands, each presenting different fire risks. Australia's forested areas are diverse and widespread, covering a wide range of ecosystems, such as tropical and subtropical rainforests, eucalypt forests, acacia forests, melaleuca forests, and casuarina forests \cite{webb1959physiognomic}. Eucalypt forests are the most extensive, spanning large areas across the country's eastern, southern, and southwestern parts \cite{miller2017fire}. These forests' thick bark and ability to resprout after fire events significantly influence fire behaviour in Australia. Particularly prone to intense fires due to the high oil content in the leaves, which increases flammability.  Tropical rainforests are found mainly in Queensland, while acacia forests are prevalent in the arid and semi-arid regions \cite{DepartmentOfAgriculture2021}.

\begin{figure}[ht!] 
\centering
\includegraphics[width=3.5in]{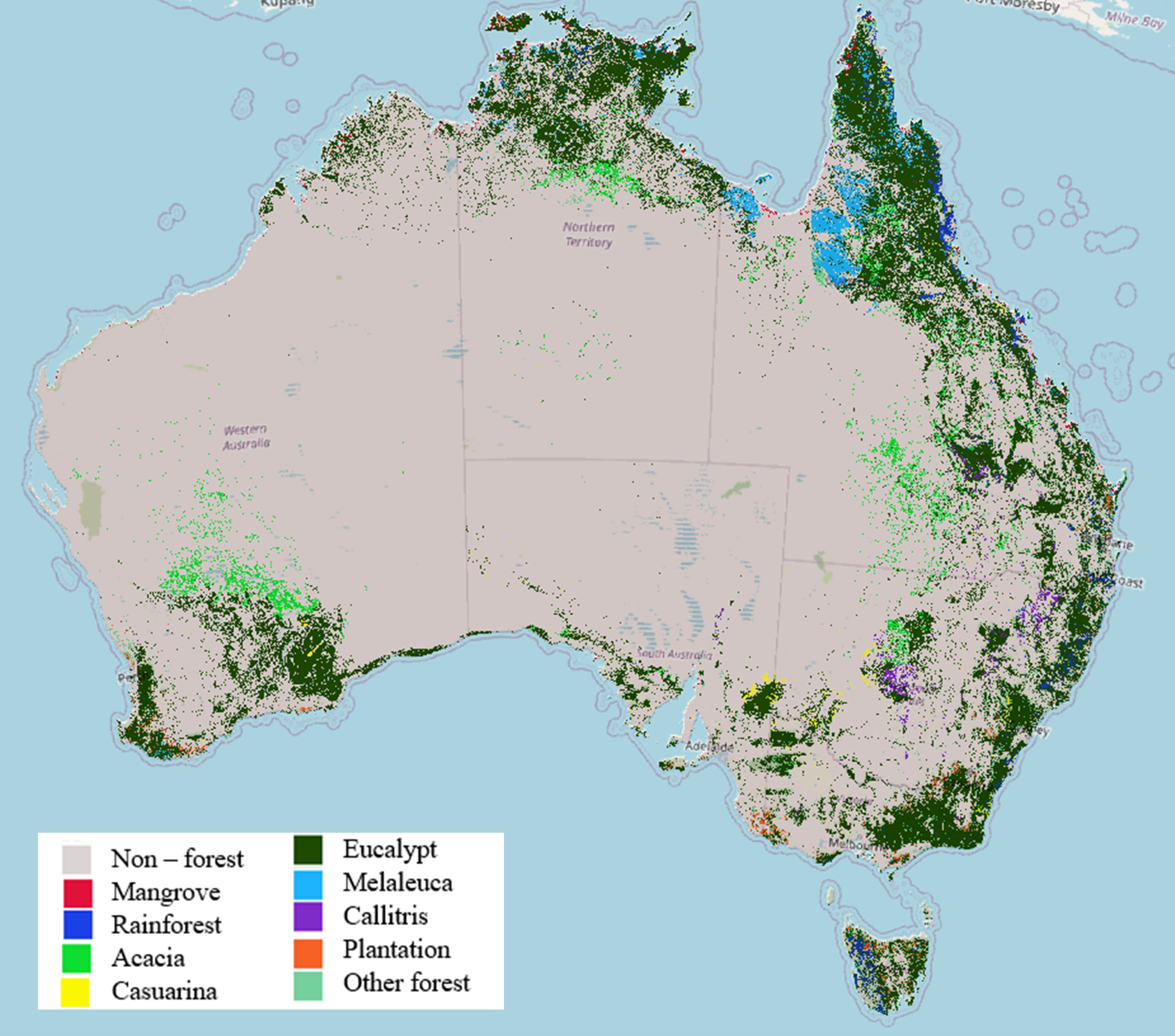}
\caption{Vegetation types across Australia (visualized on QGIS) illustrate the distribution of various forest and non-forest areas. Key vegetation types include eucalypt, rainforest, acacia, and melaleuca forests, among others. Each contributes differently to fire behaviour and severity.}
\label{figure6}
\end{figure}

The map in Fig. \ref{figure6} visually represents the distribution of different vegetation types across Australia, highlighting the diversity and spread of forested areas. Eucalypt forests, marked in dark green, are the most prevalent and are distributed extensively across the eastern, southern, and southwestern regions. These forests play a crucial role in Australia's fire dynamics due to their flammable nature, which is influenced by the high oil content in eucalypt leaves. Rainforests, in blue, are concentrated in Queensland and represent a different fire risk profile due to their typically wetter environment. Acacia forests in the arid and semi-arid regions are marked in yellow and contribute to the fire behaviour in these drier landscapes. 

The availability of fuel, such as dry vegetation, can exacerbate fire intensity, leading to rapid fire spread and increased difficulty in containment. Prioritizing areas with high fuel loads and flammable vegetation is essential for preventing uncontrollable fires.

\subsubsection{Fire Severity and Historical Trends}
Fire severity is another influencing factor, which encompasses the extent of environmental and infrastructure damage and is a critical consideration in resource allocation. Historical data on fire severity can help predict future fire patterns and identify regions requiring more attention during fire events. The use of satellite data, such as NASA’s FIRMS, has enabled researchers to map fire severity over time and assess the impact on various regions. By analyzing these patterns, it is possible to identify areas that are more prone to severe fires and allocate resources accordingly. For example, regions that have experienced repeated severe fires may require pre-emptive resource allocation to mitigate the risk of future fires.

\begin{figure}[ht!] 
\centering
\includegraphics[width=3.5in]{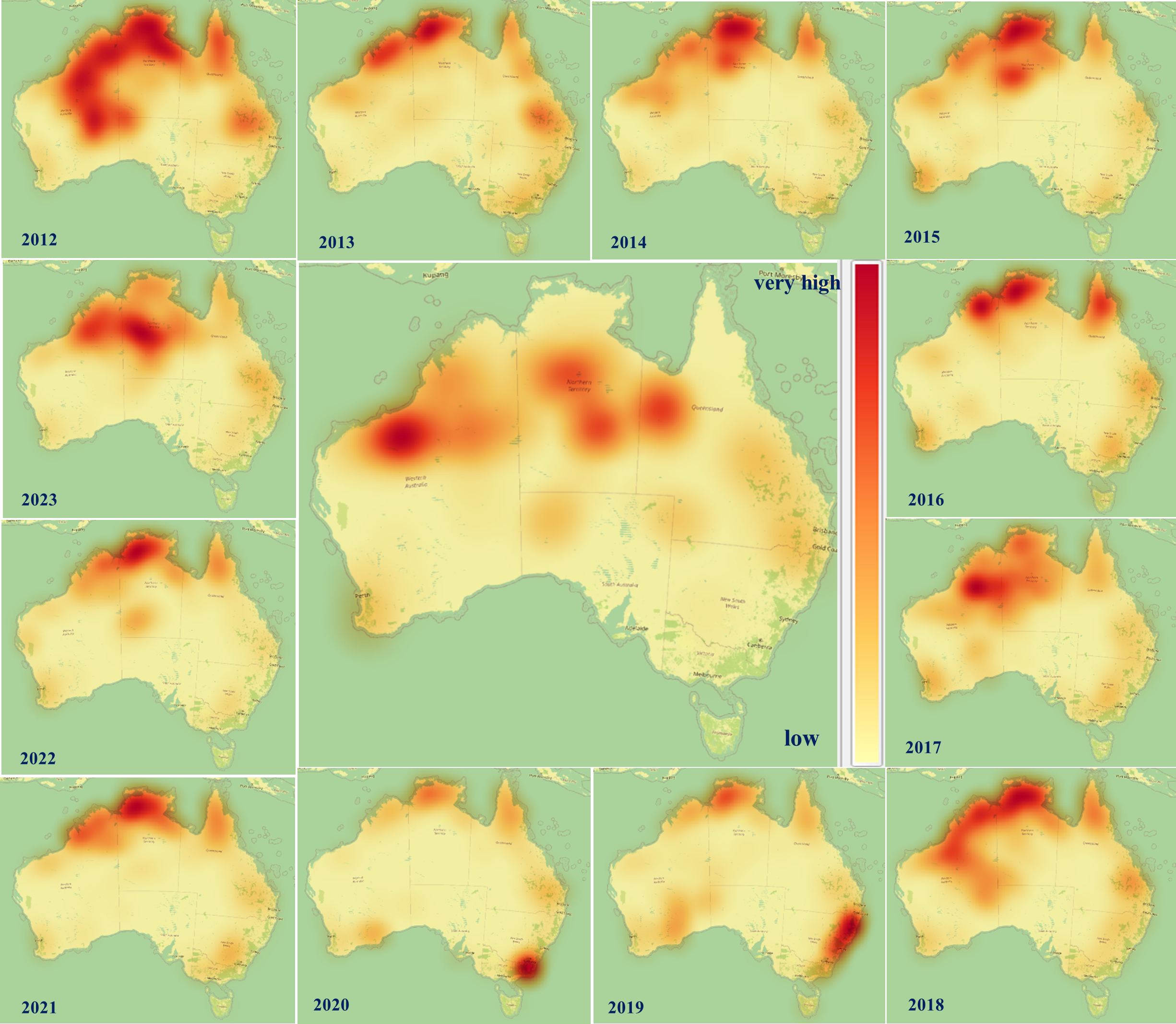}
\caption{Temporal distribution of fire severity across Australia from 2012 to 2023. The maps illustrate the intensity and spread of fire activity, with areas of high severity shown in red. The progression highlights significant fire events, particularly in northern and eastern regions, with a notable increase in intensity during certain years.}
\label{figure7}
\end{figure}

 Fig.\ref{figure7}, with a series of heatmaps visualized on QGIS, illustrates the yearly fire severity across Australia from 2012 to 2023, using data acquired from the MODIS instrument. The fire severity is categorized into four levels: low, moderate, high, and very high, represented by a color gradient from yellow to red. Areas with recurrent high fire severity, particularly in the northern and eastern regions, are evident. This visualization aids in identifying trends and patterns in fire occurrences over the 12-year period, providing valuable insights for improving fire management and mitigation strategies.

\section{Results and Discussion} 

\subsection{Fire Trend Analysis}

Understanding the temporal trends of fire events in Australia is critical for developing effective fire management strategies and preparing for future fire seasons. Australia, with its diverse climates and vegetation types, experiences varying fire conditions throughout the year, leading to fluctuating numbers of fire events. This analysis covers a comprehensive period from February 2012 to January 2024, capturing over a decade of fire activity across the continent. 

\begin{figure*}

\centering
\includegraphics[width=7in]{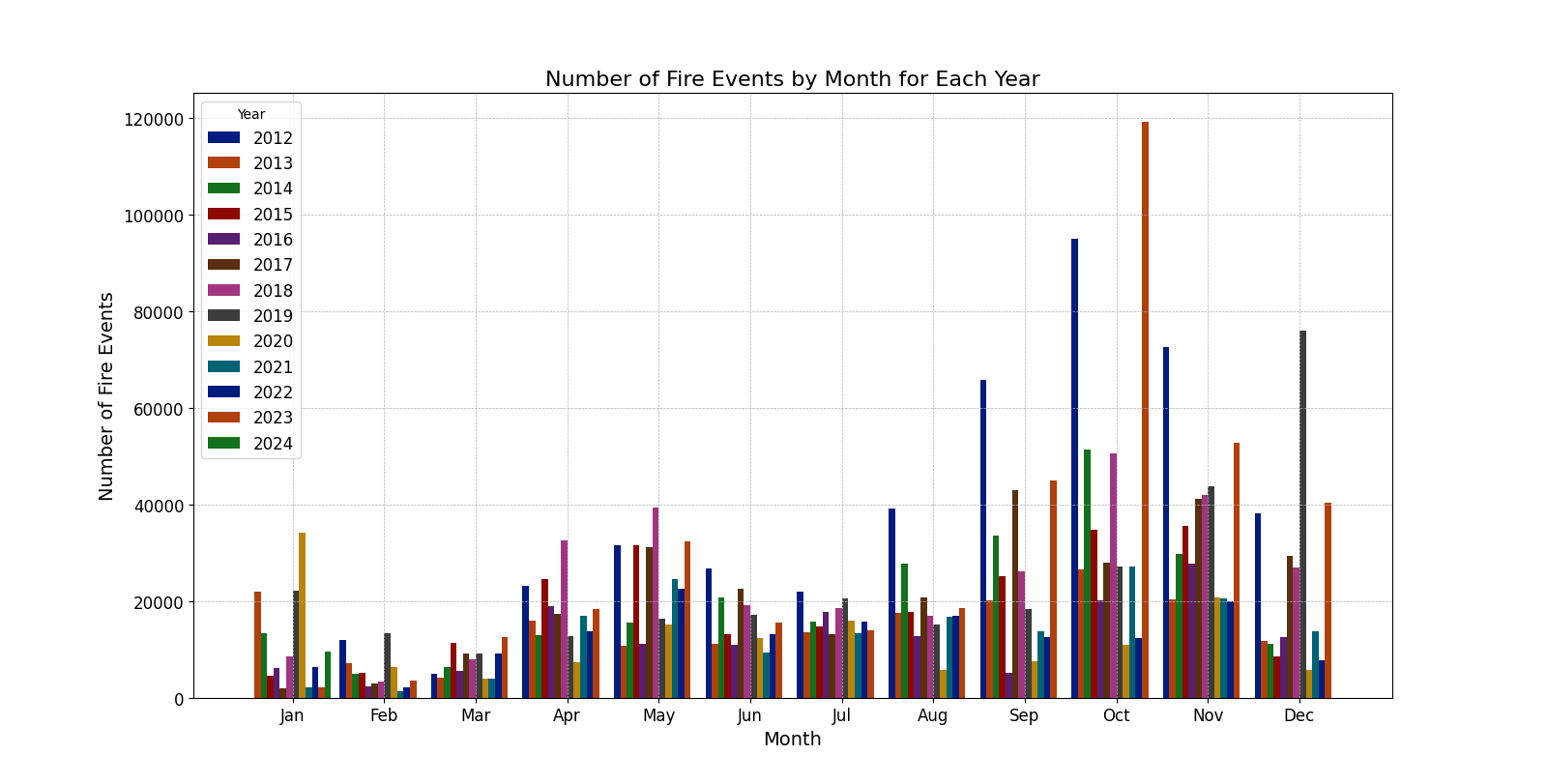}
\caption{A monthly distribution of fire events in Australia from February 2012 to January 2024 shows the increasing intensity and variability of fires over time. The data highlights significant peaks in fire activity during the spring and summer months, particularly from September to November, corresponding to Australia’s fire season.}
\label{figure8}
\end{figure*}

The bar graph (Fig. \ref{figure8}) illustrates Australia's monthly distribution of fire events from 2012 to 2024. It reveals a seasonal trend in fire occurrences, with significant peaks during spring and summer, aligning with Australia's fire season, which typically peaks from September to December. Notably, 2023 stands out with a substantial increase in fire activity, particularly in October, which recorded nearly 120,000 fire events, the highest in the observed period. 

The data reveals that fire events fluctuate annually, with certain years experiencing significantly higher activity than others. For instance, 2012 and 2023 both exhibit notable peaks in fire events, particularly in the months leading up to and during the fire season. The year 2020 also shows a high number of fire events, though slightly less than 2019, reflecting the lingering effects of the Black Summer fires. Conversely, some years, such as 2021 and 2022, show relatively lower peaks, indicating less severe fire seasons.

This trend analysis underscores the need for proactive measures to mitigate the impact of future fire seasons. The significant variability in fire event frequency from year to year, as the data reveals, suggests that fire risk is influenced by a complex interplay of factors, including weather patterns, vegetation conditions, and climate change. Certain months consistently pose higher risks, necessitating the allocation of additional firefighting resources during these times. Moreover, the extreme fire events observed in recent years, particularly the unprecedented 2023 surge, highlight the growing need for proactive measures, particularly as climate change continues influencing Australia's fire regimes.

In addition to the monthly distribution of fire counts, the Fire Severity Index (FSI) is a crucial metric for assessing fire events' intensity and potential impact \cite{edwards2013spectral}. It is calculated based on factors, including the area burned, fire intensity, and the duration of fire events within a specific timeframe. The FSI provides a standardized way to quantify and compare fire severity across regions and time periods. A higher FSI value indicates more severe fire conditions, often corresponding to increased damage and a more significant threat to life and property.

Our approach to calculating the FSI (Eq. \eqref{eq:fsi}) involves a logical construction based on standard practices \cite{laneve2020daily} in fire severity analysis. This method has been rigorously applied to ensure the reliability of our results.

\begin{equation}
\text{FSI} = \sum \left(\text{Burned Area} \times \text{Fire Intensity} \times \text{Duration}\right)
\label{eq:fsi}
\end{equation}

The formula we use integrates the critical aspects of fire dynamics, such as area burned, fire intensity, and fire event duration, to yield a single value that comprehensively represents the severity of the fire for a given year.

\begin{figure*}
\centering
\includegraphics[width=5.5in]{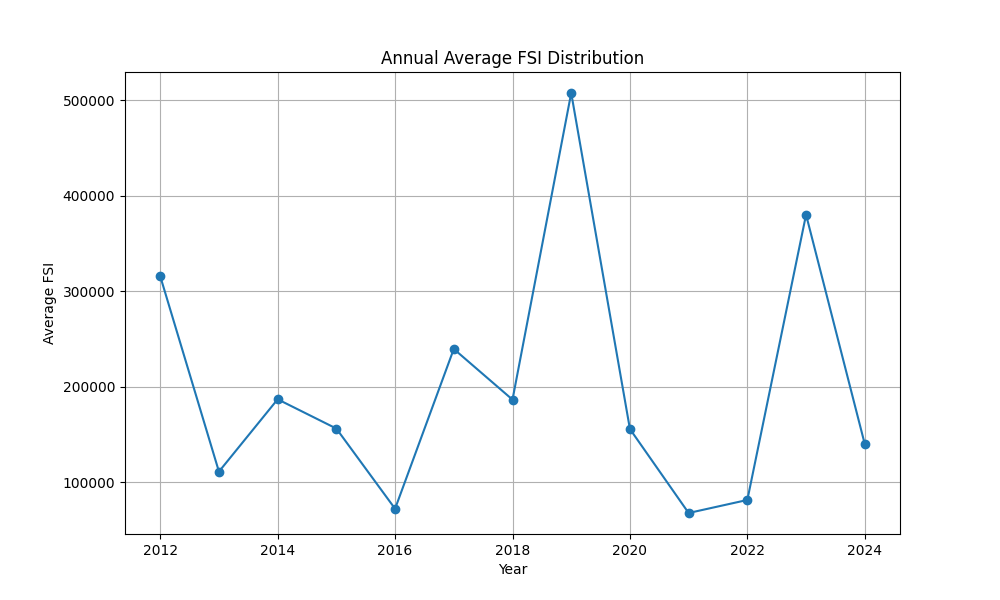}
\caption{The Annual Average FSI Distribution from 2012 to 2024 highlights the variability in fire severity across Australia. Peaks in FSI correlate with significant fire events, particularly in 2012, 2019, and 2023, reflecting the intensity and impact of these fire seasons.}
\label{figure9}
\end{figure*}

The line graph (Fig. \ref{figure9}) illustrates the annual average FSI distribution from 2012 to 2024. The graph shows significant variability in FSI values across the years, reflecting the changing fire conditions in Australia. The year 2012 witnessed one of the highest FSI values, reflecting the severe fire conditions prevalent during that year. This peak corresponds with an increase in fire events during the spring and summer months, as indicated in the monthly fire event distribution bar graph. A noticeable dip in FSI was observed in 2016, indicating a less intense fire season. This corresponds to fewer fire events and reduced fire intensity during that year, as seen in the bar graph.

The most significant peak in FSI occurred in 2019, coinciding with the ``Black Summer" fires. This period saw unprecedented fire intensity and widespread damage across Australia. The FSI analysis correlates closely with the monthly fire event distribution. For example, the high FSI values in 2019 and 2023 align with the significant peaks in fire events during these years, particularly in October and November. This indicates that not only were there more fires during these months, but the fires were also more intense and severe, contributing to the higher FSI. Conversely, years with lower FSI values, such as 2016, show fewer and less intense fire events.

\subsection{Predictive Model Evaluation and Results}

The dataset used for the predictive modelling covers a substantial period from 2012 to 2023, focusing on multiple bands and evaluating the NBR across different years. The XGBoost model employed in this study demonstrated remarkable performance in predicting fire severity using dNBR as the target variable. The Table \ref{tab:model_summary} summarizes the performance of the prediction model and the optimal hyperparameters identified during the tuning process. The model achieved an impressively low MSE of 0.0005 and a high R² score of 0.8613, showcasing its strong predictive accuracy. The low MSE indicates that the predicted dNBR values closely match the actual values, with minimal errors, while the R² score reveals that the model can explain 86.13\% of the variability in fire severity. This level of performance is highly commendable for a regression model and highlights the model's robustness. Such results suggest that the combination of spectral, topographical, and climatic features used as input data were highly effective in capturing the complex factors that contribute to fire severity, enabling the model to make accurate predictions. As a result of these metrics, the model was shown to be resilient to generalization, demonstrating that it is capable of being applied to real-world prediction scenarios of fire severity.

\begin{table*}[h!]
\centering
\caption{Prediction model performance Summary and best hyperparameters for XGBoost Model}
\resizebox{\textwidth}{!}{%
\begin{tabular}{|l|c|c|c|c|c|c|c|}
\hline
\multirow{2}{*}{\textbf{Model}} & \multirow{2}{*}{\textbf{Accuracy}} & \multirow{2}{*}{\textbf{MSE}} & \multicolumn{5}{c|}{\textbf{Hyperparameters}} \\ \cline{4-8}
                                &                                    &                               & \textbf{Boosting rounds} & \textbf{Learning Rate} & \textbf{Max Tree Depth} & \textbf{Feature Sample Rate} & \textbf{Sub-Sample Rate} \\ 
                                &                                    &                               & \textbf{(n\_estimators)}  &      \textbf{(learning\_rate)}                  & \textbf{(max\_depth)}    & \textbf{(colsample)}     & \textbf{(subsample)}     \\ \hline
XGBoost                         & 86.13 \%                             & 0.0005                        & 500                      & 0.1                    & 5                       & 0.8                      & 0.8                      \\ \hline
\end{tabular}%
}
\label{tab:model_summary}
\end{table*}

\begin{figure*}
\centering
\includegraphics[width=5in]{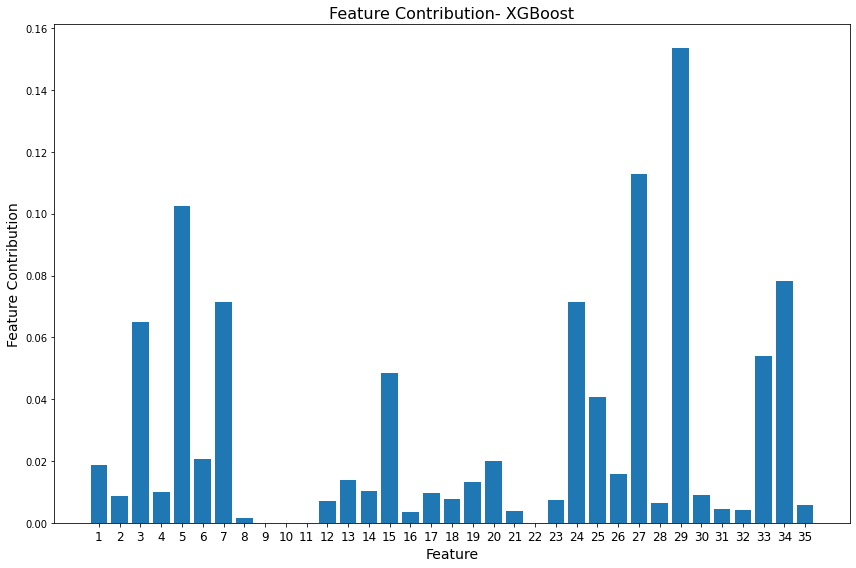}
\caption{Feature Contribution to XGBoost Model Performance: Bar plot showing the relative contribution of each feature in predicting the target variable using XGBoost. Higher values indicate more influential features in the model.}
\label{figure10}
\end{figure*}

The feature importance plot (Fig. \ref{figure10}) provides insights into which features contributed the most to the model's predictions. In XGBoost, feature importance is calculated based on how often a feature is used for splitting the data and how much it improves the prediction by reducing impurity in the decision trees. In this case, several features, particularly those with an importance score above 0.1, were identified as significant contributors. The plot shows that certain spectral bands and indices stand out as critical in predicting fire severity, while others contribute less or have no impact. The band with the highest importance is SWIR2 (Shortwave Infrared 2, Band 29), which significantly assesses burn severity and vegetation damage. This makes sense, as the SWIR2 band is sensitive to moisture content in soil and vegetation. Changes in this band often indicate the presence or absence of water, which is critical in determining how severely fire impacts an area. Areas with less moisture are likely to experience more intense fires, hence the high importance of SWIR2 in predicting fire severity.

Other bands that show considerable importance include the Burn Index (BI, Band 27), NBR (Normalized Burn Ratio, Band 24), and EVI (Enhanced Vegetation Index, Band 25). These indices are designed specifically to assess fire and vegetation health, making them intuitive choices for a model predicting fire severity. SR\_B5 (Near-Infrared, Band 5) and SR\_B7 (Shortwave Infrared 1, Band 7) also contribute significantly, reinforcing the idea that near-infrared and shortwave infrared bands, which are sensitive to vegetation health and water content, are key predictors of fire damage. Similarly, Total Precipitation (Band 34) and Mean Temperature (Band 33) have high importance in the model, which aligns with our understanding that climatic factors like temperature and precipitation are critical in determining fire risk and severity. These factors influence the dryness of vegetation and soil, contributing to the overall fire environment.

Conversely, some bands such as SR\_QA\_AEROSOL (Band 8), SR\_ATMOS\_OPACITY (Band9), SR\_CLOUD\_QA (Band 10), ST\_B6 (Thermal Band, Band 11), and QA\_RADSAT (Band 22) show no importance in the model. This can be attributed to the fact that these quality assessment and radiometric bands are generally used to ensure the integrity of the data rather than directly contribute to fire severity predictions. Their lack of importance is expected since the primary focus of the model is on spectral indices and bands directly related to vegetation, moisture content, and climatic variables rather than quality control information. This design choice ensures that the model is focused on the most relevant features for predicting fire severity.

\begin{figure*}
\centering
\includegraphics[width=5in]{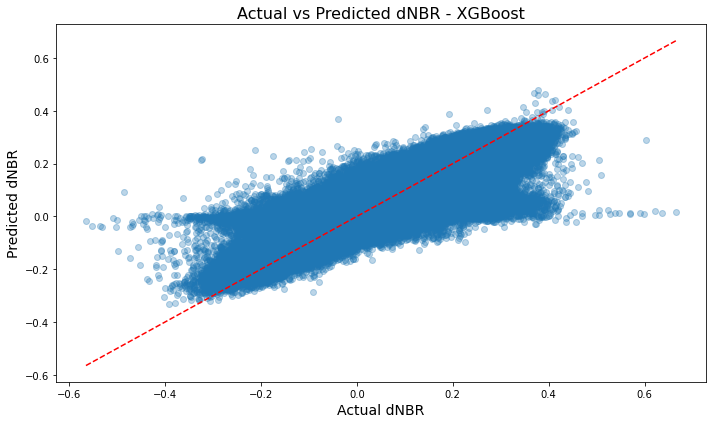}
\caption{The distribution of median NBR values across different years, providing a visual representation of the data's density and range. }
\label{figure11}
\end{figure*}

We have visualized the Actual vs. Predicted dNBR plot (Fig. \ref{figure11}), which visually compares the model’s predictions against the actual dNBR values in the test set. The data points are closely aligned with the red dashed line, which represents perfect predictions. This close alignment suggests the model performs well across a wide range of fire severity values, accurately capturing the relationship between the input features and the dNBR. Although there are a few deviations from the line, these may be attributed to factors not fully captured by the model. The plot indicates a strong correspondence between predicted and actual values, demonstrating that the model generalizes well and does not suffer from overfitting. Furthermore, the model captures the linear and non-linear relationships between the features and the target variable, effectively modelling the complexity of fire severity without a strong pattern of deviation.

\begin{figure*}
\centering
\includegraphics[width=5in]{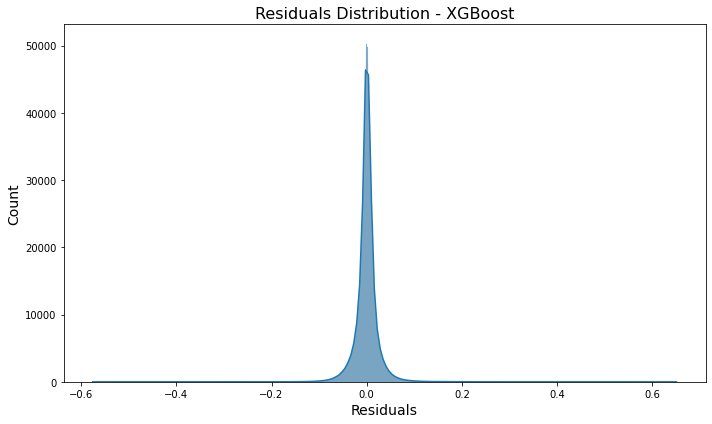}
\caption{The residuals of the predicted versus actual NBR values, highlighting the difference between the model's predictions and the actual data. }
\label{figure12}
\end{figure*}

An evaluation of a predictive model's performance using a residual plot is a critical part of regression analysis. The residuals are the differences between the observed actual values and the predicted values made by the model. The plot visually represents these residuals, typically by plotting them on the vertical axis against either the predicted values or another variable on the horizontal axis. The residual distribution (Fig. \ref{figure12}) for our model has a sharp peek around the centre, suggesting that most predictions are highly accurate, with only minor deviations. The symmetrical nature of the residual distribution and the narrow spread indicate that the model is unbiased and that errors are minimal. A well-behaved residual distribution like this strongly indicates the model’s reliability and generalization capacity.

\begin{figure*}
    
\centering
\includegraphics[width=5in]{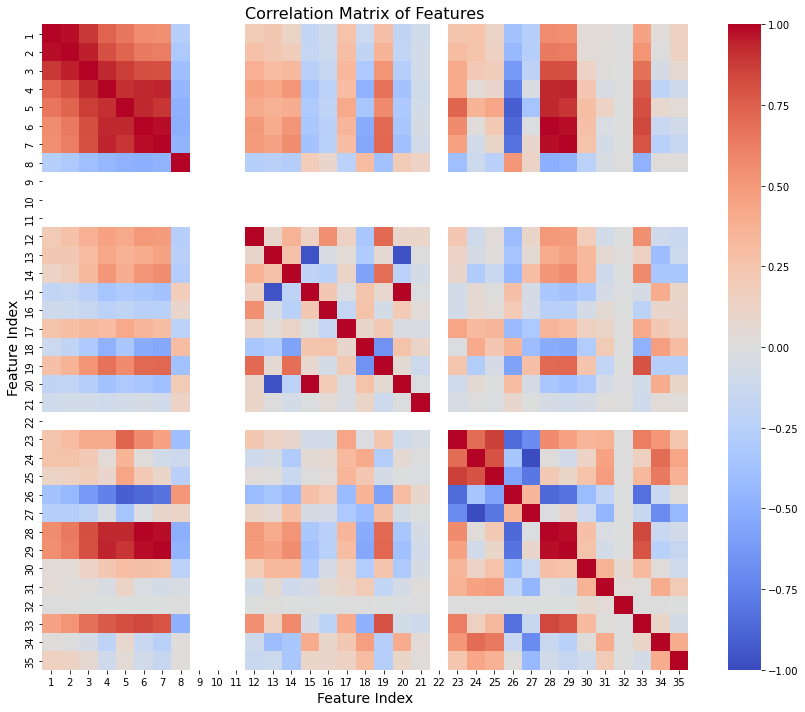}
\caption{Correlation matrix showing the relationships between the spectral bands and indices. Positive correlations are highlighted in red, while negative correlations are shown in blue.}
\label{figure13}
\end{figure*}

Understanding the correlations is essential for interpreting how features interact and contribute to the model's overall accuracy. The correlation matrix (Fig. \ref{figure13}) for our model provides valuable insights into the relationships between the various spectral, topographical, and climatic features used in the fire severity prediction model. Several notable patterns emerge from this matrix, with strong correlations between key spectral bands and indices, while others show weaker or negligible relationships.

One of the strongest correlations is between SR\_B5 and NBR. This is expected, as NBR is derived from near-infrared and shortwave infrared bands, and SR\_B5 directly influences its calculation. Both bands are highly sensitive to vegetation health and burn severity, making them crucial for predicting fire impact. Another significant relationship is observed between SWIR1 and the BI, as BI is calculated using shortwave infrared bands. This correlation indicates that SWIR1 is an effective measure of fire-related damage, as it captures variations in moisture content, which tends to decrease in areas affected by fire. Additionally, NDVI and EVI show a strong correlation, which is expected since both indices measure vegetation health and rely on similar spectral bands. Their close relationship confirms that they capture similar ecological properties, even though their calculations differ slightly.

Climatic variables such as Mean Temperature and Total Precipitation display a moderate negative correlation. This inverse relationship aligns with environmental observations in fire-prone areas, where higher temperatures are often associated with lower precipitation. These two factors play a critical role in determining the likelihood and severity of fires, as dry and hot conditions increase the risk of fire spread. Other bands, such as SWIR2, also exhibit moderate correlations with indices like NBR, further emphasizing the importance of infrared bands in capturing moisture levels and fire severity. On the other hand, several features show weak or no correlation with key predictors of fire severity. For example, QA bands such as SR\_QA\_AEROSOL, SR\_CLOUD\_QA, QA\_PIXEL, and QA\_RADSAT exhibit minimal correlation with other features. This suggests that these quality assurance bands serve their intended role in maintaining data integrity but do not directly influence fire severity predictions. Similarly, SR\_ATMOS\_OPACITY has a weak correlation with most features, indicating that atmospheric effects, as measured by this band, have little impact on the spectral data used for fire severity analysis.

Topographical features such as Slope and TPI also show limited correlations with the primary spectral indices. This suggests that while slope and terrain position may influence fire spread, they operate more independently from spectral reflectance in this dataset. Although these features are essential in understanding fire behaviour, they do not interact strongly with the spectral bands, which are more directly related to vegetation and moisture content. One interesting observation is the weak correlation of Elevation with most other features. This implies that Elevation alone may not be a significant driver of fire severity in this particular dataset. However, Elevation may still interact with other topographical factors to influence fire dynamics in specific areas. Another notable feature is ST\_DRAD, which shows moderate correlations with thermal bands like ST\_B10 and ST\_TRAD. This relationship is expected, as these bands measure atmospheric and ground-level radiation, both of which can affect fire intensity.

\subsubsection{Observations and Future Directions}

While our model demonstrates high accuracy in predicting bushfire severity based on current vegetation patterns, it is important to acknowledge the potential limitations when applying this model to future scenarios. Climate change, land use alterations, and evolving fire regimes may lead to significant shifts in vegetation composition and structure over time, which could impact the long-term reliability of our fire severity predictions. As the rate of vegetation change accelerates due to climate pressures, the historical data used in our model may become less representative of future conditions, necessitating updates or retraining.

To address these concerns, we propose the integration of a sensitivity analysis framework in future studies. This approach would allow us to explore how hypothetical changes in vegetation types and cover affect fire severity predictions and provide a more robust assessment of the model's long-term applicability. The following components outline the proposed framework:

\textbf{Vegetation Change Scenarios:}
Develop a set of plausible vegetation change scenarios based on climate projections and ecological models to reflect potential shifts in vegetation over the coming decades. This could include:
\begin{enumerate}
    \item No change (baseline): Assumes vegetation composition and cover remain stable.
    \item Moderate change: Simulates a 10\% shift in dominant species or vegetation types.
    \item Significant change: Models a more extreme 30\% shift in vegetation types, such as replacing forested areas with scrubland or grassland due to prolonged drought.
\end{enumerate}

\textbf{Model Sensitivity Testing:}
Apply our XGBoost prediction model to these hypothetical vegetation change scenarios to assess how alterations in vegetation parameters affect fire severity predictions. This step would help quantify how sensitive the model is to shifts in vegetation and identify thresholds where the model's predictive accuracy begins to degrade.

\textbf{Uncertainty Quantification:}
Calculate the range of prediction outcomes across the various vegetation scenarios to quantify the uncertainty introduced by potential future vegetation changes. It will provide insight into how the model's predictions may differ depending on the environment.

\textbf{Temporal Analysis:}
Conduct the sensitivity analysis at multiple future time horizons, such as 10, 20, and 50 years. Based on the impact of vegetation changes on fire severity predictions, we will be able to determine how often the model needs to be retrained or updated.

\textbf{Integration with Climate Models:}
Examine how both climatic and ecological shifts influence future fire severity patterns by integrating vegetation change scenarios with projected climate changes (e.g., increased temperatures and altered precipitation patterns).

This sensitivity analysis framework would allow for a more thorough evaluation of the model's long-term robustness without requiring immediate field experiments. It also provides a valuable method for understanding how the model performs under potential future conditions, guiding predictive efforts and resource planning. While this study provides a strong foundation for using historical data and machine learning to predict fire severity, it is important to recognise that the accelerating effects of climate change could result in significant deviations from historical patterns. Changes in vegetation cover, species composition, and ecosystem dynamics will all likely influence fire behaviour in ways that current models may not fully capture. As a result, the model's future accuracy depends on its ability to adapt to these changes through regular retraining and updates.

We recommend that future studies incorporate dynamic vegetation modelling, such as continuous monitoring of vegetation changes in fire-prone areas, and conduct field validations to further refine this approach. Additionally, integrating real-time satellite data and ongoing climate projections into the model would enhance its capacity to provide accurate fire severity predictions in an era of unprecedented environmental change. The model can remain relevant and effective for long-term fire risk assessments and mitigation strategies by incorporating these steps.

\subsection{Recommendations}

Based on the analysis and results obtained from the predictive model, several key recommendations are proposed to mitigate future bushfire risks in Australia. The Table \ref{tab:resource_allocation_priority}  highlights cities across the country by evaluating population density, vegetation cover, historical fire severity, and predicted fire severity to provide a comprehensive framework for resource allocation and fire prevention strategies. High-risk regions, particularly major urban centres such as Sydney, Melbourne, Brisbane, and Perth, are identified as having both dense populations and high vegetation cover, making them susceptible to severe bushfires in the future, demanding significant resource allocation for bushfire prevention and preparedness.

Other cities, such as Canberra and Darwin, also show high predicted fire severity, driven by their proximity to extensive vegetation cover and relatively high population density. Darwin, with 31\% vegetation cover, has historically experienced very high fire severity and is expected to continue facing severe fire risks, necessitating targeted fire mitigation efforts. Regions like Cairns and Broome, while less populated, show significant predicted fire severity due to their dense vegetation cover (58\% and 34\%, respectively). These areas should not be overlooked in the high allocation of firefighting resources, especially given their susceptibility to large-scale bushfires driven by environmental factors. By contrast, cities such as Hobart and Adelaide show a mix of moderate to low predicted fire severity. Historically, these cities have low fire severity and predicted risks warrant moderate resource allocation. Other cities, such as Rockhampton, Mackay, and Wollongong, are categorized under high priority based on fire risk profiles.

\begin{table*}[h!]
\centering
\caption{Resource Allocation Priority for Major Australian Cities}
\resizebox{\textwidth}{!}{%
\begin{tabular}{|l|c|c|c|c|c|}
\hline
\textbf{Region/City}      & \textbf{Population Density (people/km²)} & \textbf{Vegetation Cover (\%)} & \textbf{Historical Fire Severity} & \textbf{Predicted Fire Severity} & \textbf{Recommended Resource Allocation Priority} \\ \hline
\textbf{Sydney}      & 429   & 46                      & High    & Very High   & Very High          \\ \hline
\textbf{Melbourne}   & 500   & 40                      & Moderate   & High     & Very High         \\ \hline
\textbf{Brisbane}    & 155   & 38                      & Moderate   & High     & High              \\ \hline
\textbf{Perth }      & 315   & 46                      & Moderate   & High     & Very High         \\ \hline
\textbf{Canberra }   & 171   & 57                      & High     & High     & High  
\\ \hline
\textbf{Adelaide }   & 400  & 41                      & Moderate   & Moderate  & Moderate         \\ \hline

\textbf{Darwin }     & 44    & 31                      & Very High  & Very High   & High           \\ \hline
\textbf{Hobart }     & 125   & 52                      & Low       & Moderate   & Moderate         \\ \hline
\textbf{Cairns}      & 66    & 58                      & Very High   & High      & High       \\ \hline
\textbf{Gold Coast}  & 100    & 35                      & Moderate  & High  &  High                \\ \hline
\textbf{Newcastle}   & 123    & 40                      & Moderate  & High  &  High 
\\ \hline
\textbf{Townsville}  & 247    & 35                       & Moderate  & High  &  High               \\ \hline
\textbf{Alice Springs} & 160   & 25                      & High    & High    &  High               \\ \hline
\textbf{Broome}          & 47   & 34                     & High & High &  High                     \\ \hline
\textbf{Rockhampton}     & 118   & 37                 & Moderate  & High  &  High 
\\ \hline
\textbf{Launceston}      & 102  & 45                  & Moderate  & Moderate  &  Moderate    
\\ \hline
\textbf{Mackay}          & 95   & 42                   & Moderate  & High  &  High            
\\ \hline
\textbf{Port Hedland}          & 68    & 36            & High  & Very High   & High                                     \\ \hline
\textbf{Wollongong}          & 290   & 38                & Moderate  & High  &  High                    \\ \hline

\end{tabular}%
}
\label{tab:resource_allocation_priority}
\end{table*}

Enhanced vegetation management is also crucial in these areas. Regular practices such as controlled burns and vegetation clearance should be implemented to reduce the fuel load and mitigate the severity of potential fires. For example, regions like the Blue Mountains in New South Wales, known for their dense forests, should be regularly monitored and managed to prevent the escalation of fire severity. Strategic resource allocation, like firefighting resources, including personnel, equipment, and aerial capabilities, should be deployed based on the predicted fire severity and population risk factors. Areas with a history of severe fires and close proximity to urban centres should receive a higher allocation of resources during the fire season to enhance preparedness and response efforts.

In addition, long-term monitoring and predictive modelling should be continuously improved. Integrating real-time satellite data into predictive models will enhance the accuracy of fire severity forecasts, allowing for timely interventions. Establishing a national database that combines satellite data with local vegetation and weather information could further improve the effectiveness of predictive modelling efforts, providing a robust tool for managing future fire risks. The recommendations emphasize the importance of addressing population density and vegetation cover when planning for future bushfire prevention. The results of this study provide a solid foundation for planning controlled burns and other preventative strategies in high-risk areas. Ultimately, these strategies aim to reduce the impact of future bushfire events and protect lives and infrastructure.


\section{Conclusion}

This study provides a detailed assessment of bushfire severity mapping and predictive modelling in Australia over the last 12 years, utilizing advanced remote sensing and machine learning techniques. The results highlight the complex relationship between environmental, climatic, and topographical factors in determining fire severity. By integrating Satellite imagery using XGBoost predictive model, this research has demonstrated that spectral indices like SWIR2, NBR, and Burn Index, alongside climatic data like precipitation and temperature, are critical in forecasting future fire trends. The model's 86.13\% performance emphasizes its robustness in predicting fire severity across diverse Australian landscapes.

In addition, we mapped and analyzed bushfire severity across Australia over the past twelve years using NASA's FIRMS satellite data. Our research highlighted significant temporal and spatial patterns in fire severity, with certain regions, such as New South Wales, Victoria, and Queensland, showing higher tendencies for severe fires. The historical data analysis and future fire predictions offer valuable insights for improving fire management strategies. High-risk regions, particularly urban centres like Sydney, Melbourne, and Brisbane, where population density intersects with vulnerable vegetation cover, demand immediate attention for resource allocation. 

This paper recommends focusing on regions with dense populations and vegetation, implementing controlled burns, and monitoring fuel loads to reduce fire risks. Furthermore, the integration of real-time satellite data with predictive models could significantly enhance Australia's preparedness for future bushfire seasons. Continued advancements in fire severity modelling will be vital for mitigating the adverse effects of bushfires, protecting lives, and ensuring the sustainability of Australia's ecosystems. In our future research, we propose developing and integrating a swarm coordination model for UAVs to further enhance bushfire prediction and management in very high-risk areas identified in this study. This model would leverage real-time data from UAVs to monitor fire dynamics, offering a rapid response capability that can significantly improve the effectiveness of firefighting efforts and reduce the adverse impacts of bushfires.




\bibliographystyle{IEEEtran}
\bibliography{IEEEabrv,biblio_traps_dynamics}

\begin{thebibliography}{10}
\providecommand{\url}[1]{#1}
\csname url@samestyle\endcsname
\providecommand{\newblock}{\relax}
\providecommand{\bibinfo}[2]{#2}
\providecommand{\BIBentrySTDinterwordspacing}{\spaceskip=0pt\relax}
\providecommand{\BIBentryALTinterwordstretchfactor}{4}
\providecommand{\BIBentryALTinterwordspacing}{\spaceskip=\fontdimen2\font plus
\BIBentryALTinterwordstretchfactor\fontdimen3\font minus \fontdimen4\font\relax}
\providecommand{\BIBforeignlanguage}[2]{{%
\expandafter\ifx\csname l@#1\endcsname\relax
\typeout{** WARNING: IEEEtran.bst: No hyphenation pattern has been}%
\typeout{** loaded for the language `#1'. Using the pattern for}%
\typeout{** the default language instead.}%
\else
\language=\csname l@#1\endcsname
\fi
#2}}
\providecommand{\BIBdecl}{\relax}
\BIBdecl

\bibitem{ellis2004national}
Ellis,~S., Kanowski,~P., and Whelan,~R., ``National inquiry on bushfire mitigation and management,'' Report to the Council of Australian Governments (COAG), Canberra, Australia, 2004.

\bibitem{clarke2021fire}
Clarke,~M.~F., Kelly,~L.~T., Avitabile,~S.~C., Benshemesh,~J., Callister,~K.~E., Driscoll,~D.~A., Ewin,~P., Giljohann,~K., Haslem,~A., Kenny,~S.~A. \emph{et~al.}, ``Fire and its interactions with other drivers shape a distinctive, semi-arid ‘mallee’ecosystem,'' \emph{Frontiers in Ecology and Evolution}, vol.~9, p. 647557, 2021.

\bibitem{makumbura2024spatial}
Makumbura,~R.~K., Dissanayake,~P., Gunathilake,~M.~B., Rathnayake,~N., Kantamaneni,~K., and Rathnayake,~U., ``Spatial mapping and analysis of forest fire risk areas in sri lanka--understanding environmental significance,'' \emph{Case Studies in Chemical and Environmental Engineering}, vol.~9, p. 100680, 2024.

\bibitem{recovery2020}
``Recovery collection: Australia: Black summer bushfires 2019-2020,'' \url{https://recovery.preventionweb.net/collections/recovery-collection-australia-black-summer-bushfires-2019-2020}, 2020, accessed: 2024-07-29.

\bibitem{feizizadeh2023integrated}
Feizizadeh,~B., Omarzadeh,~D., Mohammadnejad,~V., Khallaghi,~H., Sharifi,~A., and Karkarg,~B.~G., ``An integrated approach of artificial intelligence and geoinformation techniques applied to forest fire risk modeling in gachsaran, iran,'' \emph{Journal of Environmental Planning and Management}, vol.~66, no.~6, pp. 1369--1391, 2023.

\bibitem{collins2018utility}
Collins,~L., Griffioen,~P., Newell,~G., and Mellor,~A., ``The utility of random forests for wildfire severity mapping,'' \emph{Remote sensing of Environment}, vol. 216, pp. 374--384, 2018.

\bibitem{collins2020training}
Collins,~L., McCarthy,~G., Mellor,~A., Newell,~G., and Smith,~L., ``Training data requirements for fire severity mapping using landsat imagery and random forest,'' \emph{Remote Sensing of Environment}, vol. 245, p. 111839, 2020.

\bibitem{dixon2022regional}
Dixon,~D.~J., Callow,~J.~N., Duncan,~J.~M., Setterfield,~S.~A., and Pauli,~N., ``Regional-scale fire severity mapping of eucalyptus forests with the landsat archive,'' \emph{Remote Sensing of Environment}, vol. 270, p. 112863, 2022.

\bibitem{gai2011gis}
Gai,~C., Weng,~W., and Yuan,~H., ``Gis-based forest fire risk assessment and mapping,'' in \emph{2011 Fourth International Joint Conference on Computational Sciences and Optimization}.\hskip 1em plus 0.5em minus 0.4em\relax IEEE, 2011, pp. 1240--1244.

\bibitem{dowdy2018climatological}
Dowdy,~A.~J., ``Climatological variability of fire weather in australia,'' \emph{Journal of applied meteorology and climatology}, vol.~57, no.~2, pp. 221--234, 2018.

\bibitem{nolan2020causes}
Nolan,~R.~H., Boer,~M.~M., Collins,~L., Resco~de Dios,~V., Clarke,~H., Jenkins,~M., Kenny,~B., and Bradstock,~R.~A., ``Causes and consequences of eastern australia's 2019--20 season of mega-fires.'' \emph{Global change biology}, vol.~26, no.~3, 2020.

\bibitem{vigilante2024factors}
Vigilante,~T., Goonack,~C., Williams,~D., Joseph,~A., Woolley,~L.-A., and Fisher,~R., ``Factors enabling fire management outcomes in indigenous savanna fire management projects in western australia,'' \emph{International Journal of Wildland Fire}, vol.~33, no.~9, 2024.

\bibitem{partheepan2023a}
Partheepan,~S., Sanati,~F., and Hassan,~J., ``Autonomous unmanned aerial vehicles in bushfire management: Challenges and opportunities,'' \emph{Drones}, vol.~7, no.~1, p.~47, 2023.

\bibitem{partheepan2023b}
Partheepan,~S., Hassan,~J., and Sanati,~F., ``An analysis of pre-trained models versus custom deep learning models for forest fire detection,'' in \emph{2023 33rd International Telecommunication Networks and Applications Conference}.\hskip 1em plus 0.5em minus 0.4em\relax IEEE, 2023, pp. 277--282.

\bibitem{ABS2024}
\BIBentryALTinterwordspacing
{Australian Bureau of Statistics}, ``Population clock pyramid,'' 2024, accessed: 2024-09-10. [Online]. Available: \url{https://www.abs.gov.au/statistics/people/population/population-clock-pyramid}
\BIBentrySTDinterwordspacing

\bibitem{brown1997australian}
Brown,~A., Turnbull,~J., and Booth,~T., ``The australian environment,'' \emph{ACIAR MONOGRAPH SERIES}, vol.~24, pp. 1--18, 1997.

\bibitem{ramankutty2006global}
Ramankutty,~N., Graumlich,~L., Achard,~F., Alves,~D., Chhabra,~A., DeFries,~R.~S., Foley,~J.~A., Geist,~H., Houghton,~R.~A., Goldewijk,~K.~K. \emph{et~al.}, ``Global land-cover change: Recent progress, remaining challenges,'' \emph{Land-use and land-cover change: local processes and global impacts}, pp. 9--39, 2006.

\bibitem{xiong2009overview}
Xiong,~X., Wenny,~B.~N., and Barnes,~W.~D., ``Overview of nasa earth observing systems terra and aqua moderate resolution imaging spectroradiometer instrument calibration algorithms and on-orbit performance,'' \emph{Journal of Applied Remote Sensing}, vol.~3, no.~1, p. 032501, 2009.

\bibitem{giglio2003enhanced}
Giglio,~L., Descloitres,~J., Justice,~C.~O., and Kaufman,~Y.~J., ``An enhanced contextual fire detection algorithm for modis,'' \emph{Remote sensing of environment}, vol.~87, no. 2-3, pp. 273--282, 2003.

\bibitem{chuvieco2020satellite}
Chuvieco,~E., Aguado,~I., Salas,~J., Garc{\'\i}a,~M., Yebra,~M., and Oliva,~P., ``Satellite remote sensing contributions to wildland fire science and management,'' \emph{Current Forestry Reports}, vol.~6, pp. 81--96, 2020.

\bibitem{QGIS2024}
\BIBentryALTinterwordspacing
{QGIS Development Team}, ``{QGIS Geographic Information System},'' 2024, version 3.37, Accessed: 2024-09-10. [Online]. Available: \url{https://qgis.org}
\BIBentrySTDinterwordspacing

\bibitem{vitalis2020cityjson}
Vitalis,~S., Arroyo~Ohori,~K., and Stoter,~J., ``Cityjson in qgis: Development of an open-source plugin,'' \emph{Transactions in GIS}, vol.~24, no.~5, pp. 1147--1164, 2020.

\bibitem{AustraliaROI2024}
\BIBentryALTinterwordspacing
{Geoscience Australia}, ``Australia's geographic coordinates and regions of interest,'' 2024, accessed: 2024-09-10. [Online]. Available: \url{https://www.ga.gov.au/scientific-topics/national-location-information/landforms/continental-extent}
\BIBentrySTDinterwordspacing

\bibitem{rouse1974monitoring}
Rouse,~J.~W., Haas,~R.~H., Schell,~J.~A., Deering,~D.~W. \emph{et~al.}, ``Monitoring vegetation systems in the great plains with erts,'' \emph{NASA Spec. Publ}, vol. 351, no.~1, p. 309, 1974.

\bibitem{key2006landscape}
Key,~C.~H., Benson,~N.~C. \emph{et~al.}, ``Landscape assessment (la),'' \emph{FIREMON: Fire effects monitoring and inventory system}, vol. 164, pp. LA--1, 2006.

\bibitem{huete2002overview}
Huete,~A., Didan,~K., Miura,~T., Rodriguez,~E.~P., Gao,~X., and Ferreira,~L.~G., ``Overview of the radiometric and biophysical performance of the modis vegetation indices,'' \emph{Remote sensing of environment}, vol.~83, no. 1-2, pp. 195--213, 2002.

\bibitem{mcfeeters1996use}
McFeeters,~S.~K., ``The use of the normalized difference water index (ndwi) in the delineation of open water features,'' \emph{International journal of remote sensing}, vol.~17, no.~7, pp. 1425--1432, 1996.

\bibitem{chuvieco2006use}
Chuvieco,~E., Ria{\~n}o,~D., Danson,~F., and Martin,~P., ``Use of a radiative transfer model to simulate the postfire spectral response to burn severity,'' \emph{Journal of Geophysical Research: Biogeosciences}, vol. 111, no.~G4, 2006.

\bibitem{miller2007quantifying}
Miller,~J.~D. and Thode,~A.~E., ``Quantifying burn severity in a heterogeneous landscape with a relative version of the delta normalized burn ratio (dnbr),'' \emph{Remote sensing of Environment}, vol. 109, no.~1, pp. 66--80, 2007.

\bibitem{chen2016xgboost}
Chen,~T. and Guestrin,~C., ``Xgboost: A scalable tree boosting system,'' in \emph{Proceedings of the 22nd acm sigkdd international conference on knowledge discovery and data mining}, 2016, pp. 785--794.

\bibitem{worldpop_australia_2020}
\BIBentryALTinterwordspacing
{WorldPop}, ``100m population density for australia, 2020,'' Dataset, 2020, accessed via Google Earth Engine. [Online]. Available: \url{https://www.worldpop.org/}
\BIBentrySTDinterwordspacing

\bibitem{trollope1984fire}
Trollope,~W., ``Fire behaviour,'' in \emph{Ecological effects of fire in South African ecosystems}.\hskip 1em plus 0.5em minus 0.4em\relax Springer, 1984, pp. 199--217.

\bibitem{webb1959physiognomic}
Webb,~L.~J., ``A physiognomic classification of australian rain forests,'' \emph{The Journal of Ecology}, pp. 551--570, 1959.

\bibitem{miller2017fire}
Miller,~B.~P. and MurPhy,~B.~P., ``Fire and australian vegetation,'' \emph{Australian Vegetation. Cambridge University Press, Cambridge}, vol.~3, pp. 113--134, 2017.

\bibitem{DepartmentOfAgriculture2021}
{Department of Agriculture, Water and the Environment}, ``Department of agriculture, water and the environment annual report 2021-22,'' \url{https://www.transparency.gov.au/publications/agriculture-water-and-the-environment/department-of-agriculture-water-and-the-environment/department-of-agriculture-water-and-the-environment-anual-report-21-22}, 2021.

\bibitem{edwards2013spectral}
Edwards,~A.~C., Maier,~S.~W., Hutley,~L.~B., Williams,~R.~J., and Russell-Smith,~J., ``Spectral analysis of fire severity in north australian tropical savannas,'' \emph{Remote Sensing of Environment}, vol. 136, pp. 56--65, 2013.

\bibitem{laneve2020daily}
Laneve,~G., Pampanoni,~V., and Uddien~Shaik,~R., ``The daily fire hazard index: A fire danger rating method for mediterranean areas,'' \emph{Remote Sensing}, vol.~12, no.~15, p. 2356, 2020.

\end{thebibliography}

\end{document}